# A reconstruction algorithm of electrical impedance tomography based on one-dimensional convolutional neural network

Zhenzhong Song[a], Jianping Li [a,*], Jiafeng Yao[b], Linying Wang[c], Dan Zhu[c], Lvjun Zhang[c], Jianming Wen[a], Nen Wan[a], Jijie Ma[a], Yu Zhang[a] and Zengfeng Gao[d]

[a]*Key Laboratory of Intelligent Operation and Maintenance Technology & Equipment for Urban Rail Transit of Zhejiang Province, the Institute of Precision Machinery and Smart Structure, College of Engineering, Zhejiang Normal University, Jinhua, Zhejiang 321004, China*

[b]*The College of Mechanical and Electrical Engineering, Nanjing University of Aeronautics and Astronautics, Nanjing, 210016 China*

[c]*The Department of Respiration and Critical Care Medicine, Affiliated Jinhua Hospital, Zhejiang University School of Medicine, Jinhua Municipal Central Hospital, Jinhua,321000, China*

[d]*The Division of Fundamental Engineering, Department of Mechanical Engineering, Graduate School of Science and Engineering, Chiba University, Chiba 263-8522, Japan*

**Abstract**

Electrical impedance tomography (EIT) is a novel computational imaging technology. In order to improve the quality and spatial resolution of the reconstructed images, the G-CNN and HG-CNN algorithms are proposed based on a one-dimensional convolutional neural network (1D-CNN) in this paper. The input of the 1D-CNN is the reconstructed conductivity distribution obtained by the GVSPM algorithm or the H-GVSPM algorithm. The reconstructed images with higher resolution are obtained through the calculation of 1D-CNN. Finally, the Hadamard product is applied to calculate the output of the 1D-CNN. In the simulation results of the lung cross-section models, the correlation coefficients of the G-CNN algorithm and HG-CNN algorithm maximumly are 2.52 times and 2.20 times greater than the GVSPM algorithm and H-GVSPM algorithm, respectively. In the results of the simulation and experiment, the reconstructed images of the G-CNN and HG-CNN algorithms are distortion-free. In addition, the artifacts of the reconstructed images are diminished after calculations of the Hadamard product. This research provides a reference method for improving the quality of the reconstructed images so that EIT is better applied in medical detection.

* Corresponding author. Tel.: 178-0589-8098.
*E-mail address*: Lijp@zjnu.cn (J. Li).

## 1. Introduction

Medical imaging is a medical method that uses imaging techniques to obtain information about internal organizational structure and function. Medical imaging techniques mainly include X-ray, positron emission tomography (PET), computed tomography (CT), magnetic resonance imaging (MRI), and electrical impedance tomography (EIT). Here, EIT is a non-invasive diagnostic technique with the characteristic of visualization, non-invasiveness, and cheapness [1-3]. The main principle is to estimate the conductivity distribution or the conductivity change of internal tissue by obtaining the boundary voltage. Firstly, it is necessary to attach electrodes to the edge of the object's area. Subsequently, the excitation current is delivered to the area through the electrodes. Secondly, the boundary voltage is measured [4, 5]. Finally, the image reconstruction algorithm is applied to reconstruct the conductivity distribution or changes within the internal tissue. At present, EIT has achieved dynamic imaging for real-time monitoring. It is widely applied for medical imaging [6-10], material nondestructive diagnosis, and industrial detection [11, 12]. For example, Dang et al. applied EIT to detect gas-liquid two-phase flow and improved the visualization of two-phase flow utilizing the eigenvalue correlation method [13]. Ren et al. adopted EIT to diagnose concrete damage [14], [15]. Hu et al. applied breast shape deformation to create deformable EIT with various shapes. The deformable EIT is expected to enhance the imaging resolution of breast appearance [16]. In 2020, Masi et al. utilized 3D EIT scanning to monitor the internal erosion processes in porous media [17]. N. Omer et al. applied EIT technology as a new parameter for evaluating pleural effusion volume in 2021 [18]. In 2022, Sannamani et al. utilized EIT to detect damage in non-planar carbon fiber-reinforced polymers [19]. As a result, EIT shows excellent performance in biomedical applications, two-phase flow analysis, and other fields.

It illustrates that EIT has better development and application in many fields. However, poor spatial resolution will seriously affect the visual effect of the reconstructed images. It may cause misinformation so that it is impossible to obtain accurate information about the measured object. Therefore, it is essential to improve the spatial resolution of reconstructed images for electrical impedance tomography. A higher-resolution reconstructed image will be obtained by optimizing the algorithm or proposing a new algorithm. For example, Takei et al. adopted the Generalized Vector Sampled Pattern Matching (GVSPM) algorithm to produce the reconstructed images of EIT [20, 21]. Song et al. applied the Hadamard product to optimize the GVSPM algorithm and obtained a reconstructed image with higher resolution in an eight-electrode system [22]. Similarly, Liu et al. proposed a method for reconstructing images based on structure-aware sparse Bayesian learning (SA-SBL), which improved the spatial resolution of reconstructed images [23, 24]. Xu et al. proposed an algorithm TSVD+ART based on the truncated singular value decomposition (TSVD) in 2021 [25]. Adler et al. have applied neural networks to reconstruct images for EIT in a long time ago [26]. Neural networks exhibit strong nonlinear, adaptive, and self-learning characteristics. It avoids the complex analysis of calculating the sensitivity matrix. Ren et al. proposed a deep neural network called RCRC to achieve filtering, smoothing, and prediction of dynamically reconstructing conductivity [27]. In 2022, Zhan et al. proposed a high-fidelity shape reconstruction of multi-phase conductivity using deep discrete representation for EIT [28]. Zhang et al. proposed a deep neural network that consists of multi-layer, fully connected networks. The deep neural network has a superior ability for nonlinear fitting [29]. Wang et al. proposed a method for reconstructing images in EIT using a Gray Wolf optimized radial basis function neural network. It achieves a high image resolution of the reconstructed images [30]. Duran et al. also adopted a convolutional neural network (CNN) to realize reconstructing images [31-33] and other related applications. The H-GVSPM and GVSPM algorithms aim to achieve reconstructed images in an eight-electrode system with a better resolution. The H-GVSPM algorithm effectively reduces the influence of artifacts in reconstructed images. However, the resolution of the reconstructed images may not meet the requirements for clinical monitoring or other fields. There will be some distortion to some extent. Therefore, improving the resolution of reconstructed images is a crucial issue for the H-GVSPM and GVSPM algorithms. Then, it is also essential to make the reconstructed images avoid distortion.

In order to enhance the resolution and correlation coefficients of the reconstructed images obtained by the GVSPM algorithm and H-GVSPM algorithm, the G-CNN algorithm and HG-CNN algorithm are proposed for reconstructing images based on one-dimensional convolutional neural network (1D-CNN). The conductivity distribution of reconstructed images calculated by the GVSPM algorithm and the H-GVSPM algorithm is used as input for the 1D-CNN. 1D-CNN is trained under the condition that the actual electrical conductivity is the output. The higher-resolution reconstructed images are obtained through the calculation of 1D-CNN. The conductivity distribution of the reconstructed images is directly applied as the final output in the case that the GVSPM algorithm and H-GVSPM algorithm are employed to reconstruct images. If the quality of the reconstructed image is analyzed, the consistency between the reconstructed conductivity distribution and the actual conductivity distribution should be considered. It is also worth considering how to enhance the correlation between them further. If the correlation is small, the reconstructed images are prone to distortion or contain many artifacts. In this research, 1D-CNN is applied to optimize the GVSPM algorithm and the H-GVSPM algorithm to enhance the correlation. Firstly, the reconstructed images are obtained using the GVSPM algorithm or H-GVSPM algorithm, and the conductivity distribution is extracted. Secondly, the reconstructed conductivity distribution and the actual conductivity distribution are taken as training samples for the 1D-CNN. After the calculation of the 1D-CNN, the new reconstructed images and the conductivity distribution are obtained. Finally, the correlation coefficient is analyzed between the reconstructed conductivity distribution and the actual conductivity distribution. After the optimization of the 1D-CNN, the correlation coefficient increases significantly. The reconstructed images will not appear distorted.

## 2. Theoretical analysis

### *2.1. GVSPM algorithm and H-GVSPM algorithm*

The high-quality reconstructed images are going to be produced by using an exemplary algorithm. It is an essential phase in the process of reconstructing the images. Six algorithms are applied to achieve the reconstructed images in this study. Next, the quality of the reconstructed images is compared and evaluated. The GVSPM algorithm is a type of algorithm whose goal is to perform iterative operations repeatedly until the predicted value of the objective function is achieved. The expected output is closer to the real value by constantly correcting errors. In the operation process, the 2-norm of the vector is applied to normalize the sequence matrix of the boundary voltage and Jacobian matrix [20-22].

The Hadamard product is a computational rule of binary operations on two matrices with the same dimension. The resulting matrix has the same dimensions as the original matrix, and each element is the product of the corresponding elements from the two original matrices [34], [35]. The Hadamard product is applied to optimize the GVSPM algorithm to obtain the H-GVSPM algorithm. The conductivity distribution obtained by the GVSPM algorithm is calculated by using the calculation rule of the Hadamard product to derive a new conductivity distribution. The equation of H-GVSPM is expressed by the following (1):

$$\sigma_a = \sigma \circ \sigma \tag{1}$$

here, $\sigma$ represents the conductivity distribution of the reconstructed images obtained by GVSPM. $\sigma_a$ represents the conductivity distribution obtained by H-GVSPM. $\sigma$ and $\sigma_a$ are matrices with the same dimension.

### *2.2. G-CNN algorithm and HG-CNN algorithm*

1D-CNN is one of the deep learning algorithms that exhibit strong robustness, generalization ability, and deep feature extraction capability. As shown in Fig. 1, the GVSPM algorithm is applied to reconstruct the conductivity distribution of the measured area after obtaining the boundary voltage distribution. The conductivity distribution reconstructed by GVSPM is input into 1D-CNN for prediction analysis. The final output result of 1D-CNN is the final conductivity distribution. That is the G-CNN algorithm. Similarly, the HG-CNN algorithm utilizes 1D-CNN to post-process the conductivity distribution calculated by the H-GVSPM algorithm. Before training the 1D-CNN, it is necessary to extract the boundary voltage from the training samples. The samples of the boundary voltage are successively substituted into the GVSPM algorithm or H-GVSPM algorithm to obtain the preliminary conductivity distribution. Finally, the preliminary conductivity distribution and the actual conductivity distribution are combined to create the final training sample of 1D-CNN.

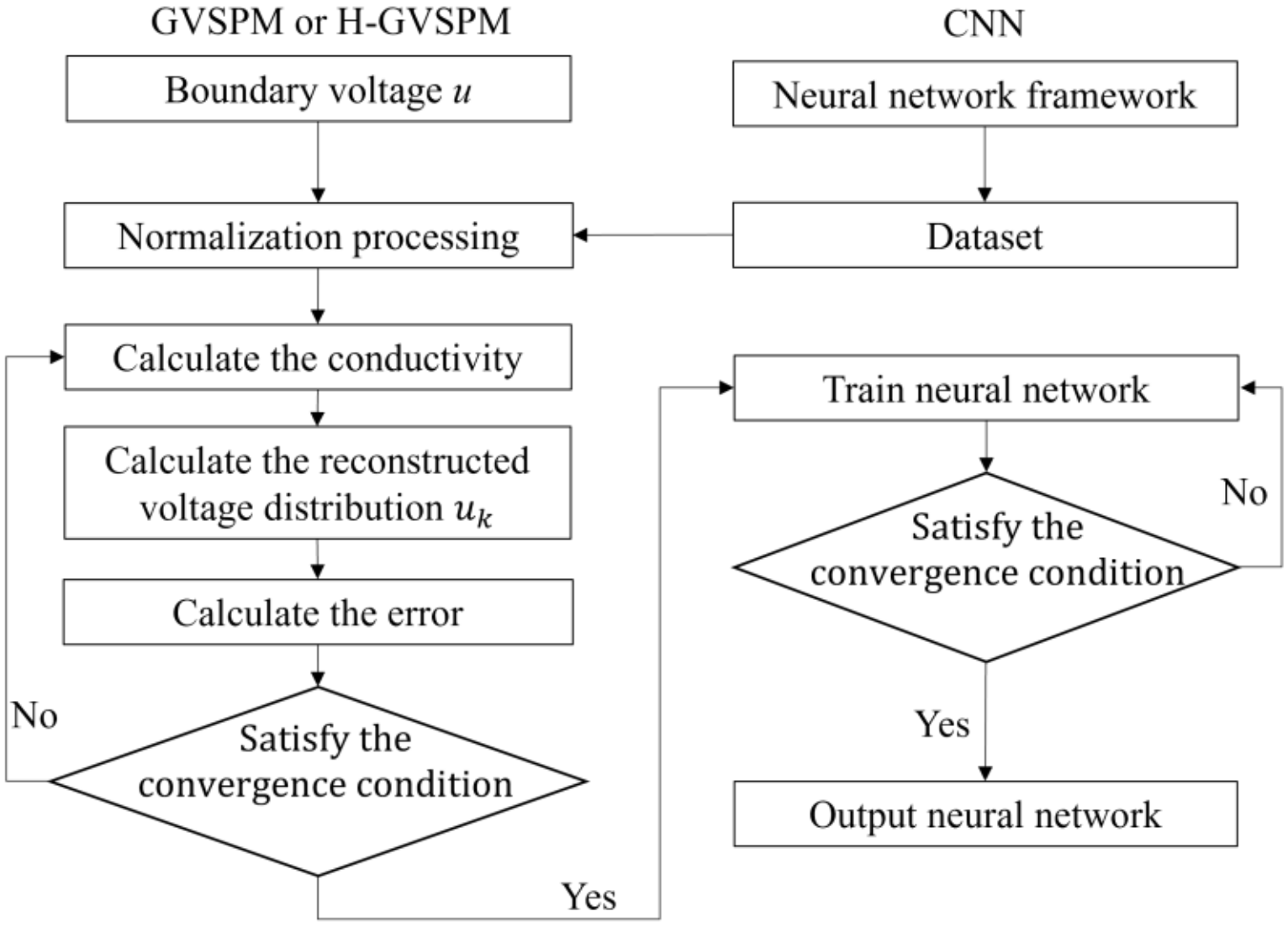

Fig. 1. The structure of G-CNN Algorithm and HG-CNN Algorithm.

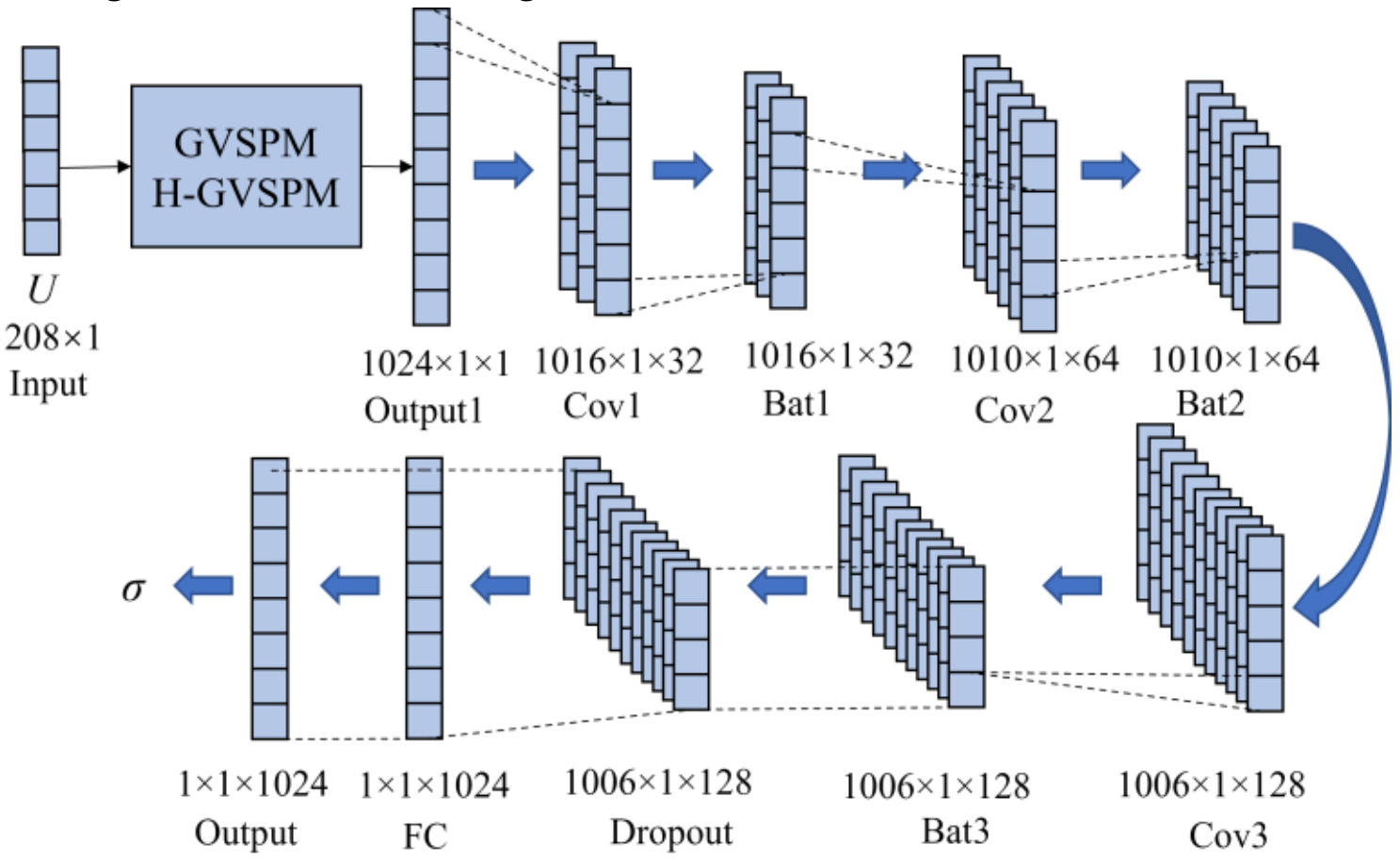

Fig. 2. The structure of the 1D-CNN. It includes three convolutional layers and three normalization layers.

The 1D-CNN involved in this study comprises three convolutional layers, three normalization layers, one fully connected layer, and one dropout layer. The structure is shown in Fig. 2. Three convolutional layers are labeled as Cov1, Cov2, and Cov3. Three normalization layers are Bat1, Bat2, and Bat3. FC refers to the fully connected layer. Since the entire measured area is divided into 1024 finite elements, the input dimension of CNN is 1024×1×1. The parameters of the convolutional kernel are 9×1×32, 7×1×64, and 5×1×128, respectively, for three convolutional layers. The step size is one. Dropout is the dropout layer where 10% of nodes are randomly eliminated. The number of nodes in the fully connected layer is 1024.

### *2.3. GH-CNN algorithm and HGH-CNN algorithm*

The Hadamard product effectively reduces artifacts [22]. Therefore, the Hadamard product is applied to calculate the conductivity obtained by the G-CNN algorithm and the HG-CNN algorithm in this study. They are named the GH-CNN algorithm and the HGH-CNN algorithm. As shown in Fig. 3, the GH-CNN algorithm and the HGH-CNN algorithm are mainly applied to explore a method for removing the artifacts of the reconstructed images.

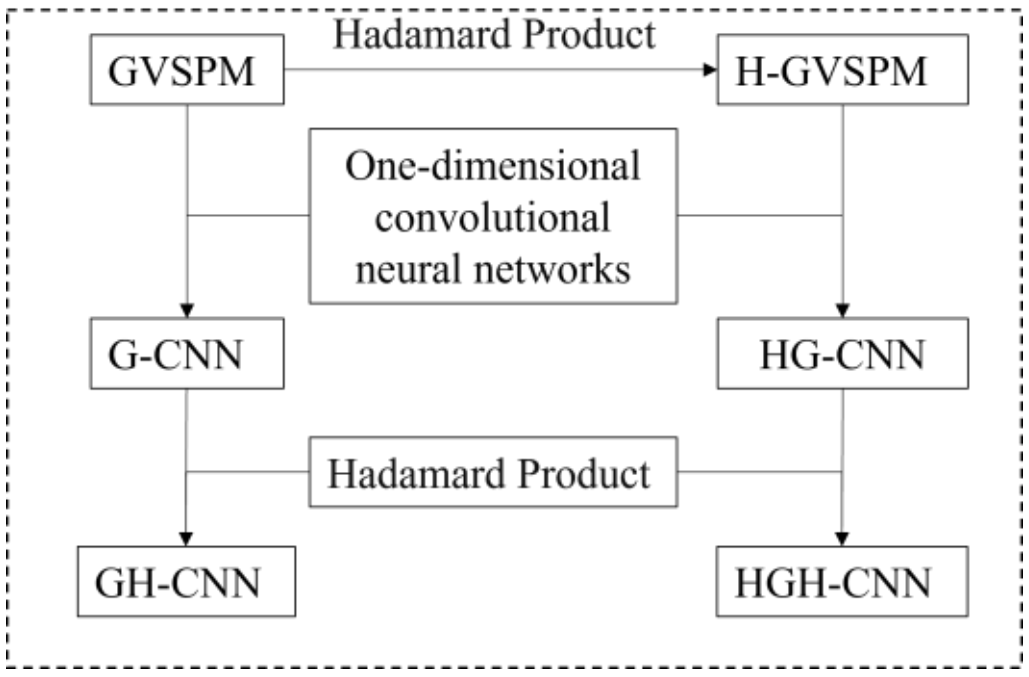

Fig. 3. The connection between each algorithm.

## 3. Simulation analysis

### *3.1. Collecting datasets*

In this study, the 1D-CNN is applied as the post-processing method for the GVSPM algorithm and the H-GVSPM algorithm. A large number of datasets need to be collected for training the 1D-CNN. The datasets include the boundary voltage and the conductivity distribution for a variety of models. The number, size, location, and conductivity values of the objects may be different in different models. As shown in Fig. 4, the lung cross-section models are divided into three models: single-target model, double-target model, and three-target model.

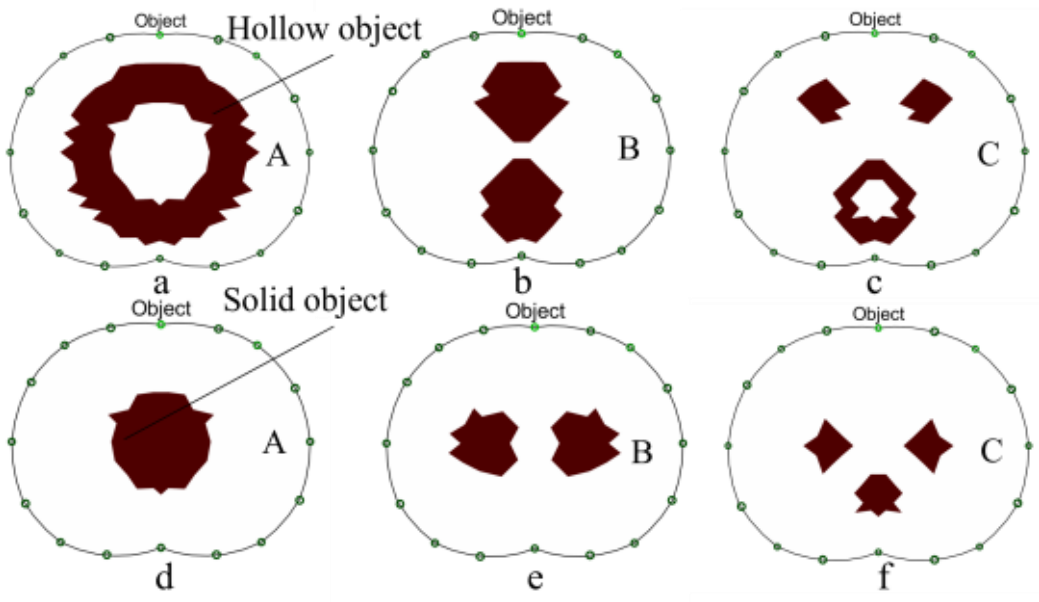

Fig. 4. Datasets models of lung cross-section models.

In Fig. 4, A represents the single-target model. B represents the double-target model, and C represents the three-target model. In the single-target model, the background conductivity is set to one. At the same time, the size of the area remains unchanged. If the sizes of the objects are different, the conductivity distribution and boundary voltage distribution in the datasets will also differ. Therefore, many datasets are generated by altering the size, position, and conductivity of the object. Similarly, a comparable approach is applied to acquire datasets for the double-target model and the three-target model. In particular, the lung cross-section models include both the solid-object models and the hollow-object models. Finally, the number of datasets for the lung cross-section models is 69,000.

### *3.2. Training one-dimensional convolutional neural networks*

The computer system applied for training the neural network is DELL Inc. in this research. The system model is Precision 7920 Tower. The processor is Intel(R)Xeon(R)Gold 6248R CPU@3.00GHz (5 CPUs) ~3.0 GHz. The operating system is Windows 10 Pro 64-bit, and the memory is 131072MB RAM. The chip type is NVIDIA GeForce RTX 3080 Ti, and the total memory is approximately 77458 MB. The display memory is 12100 MB (VRAM). In the process of training 1D-CNN, the neural network structure needs to be initialized. The maximum number of training iterations is ten. The learning rate is 0.01, and the optimizer is Adaptive Moment Estimation (ADAM).

### *3.3. Simulation results of lung cross-section models*

In order to evaluate the performance and imaging effects of the GH-CNN algorithm, HGH-CNN algorithm, G-CNN algorithm, and HG-CNN algorithm. Ten models of lung cross-section models were randomly selected in this research. On the one hand, the size and conductivity values of the objects may differ in the ten models. On the other hand, there are solid objects and hollow objects. The simulation results are shown in Fig. 5. Here, the objects from the first model to the fourth model are hollow. The size of two objects in Model III and Model VI is relatively small in all models. In the meantime, the size of a single object is relatively large in Model V. Therefore, the computer randomly selects simulation models of lung cross-sections in this study. The reconstructed images are generated for the ten selected groups of models.

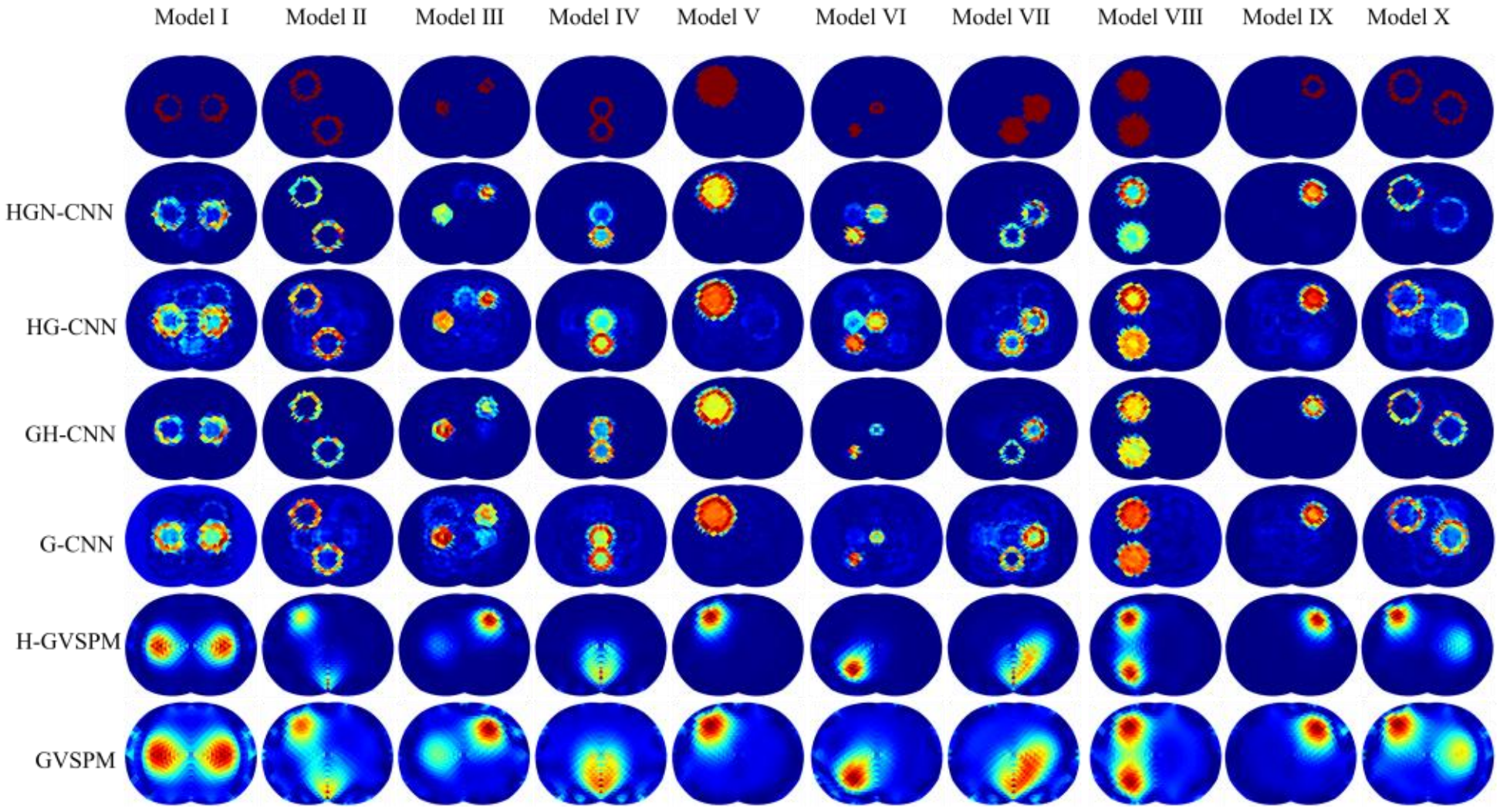

Fig. 5. The simulation results of the lung cross-section models.

As can be seen from Fig. 5, the position, size, and number of the objects are going to be roughly determined in the reconstructed images obtained by the GVSPM algorithm and H-GVSPM algorithm. The artifacts of the H-GVSPM algorithm are fewer than those of the GVSPM algorithm. However, the reconstruction effect of the H-GVSPM algorithm is not optimal. The GVSPM algorithm and H-GVSPM algorithm do not determine the internal contour in the case that there is a hollow object. The object contour of the reconstructed image is clearly visible for both the G-CNN algorithm and the HG-CNN algorithm. If two smaller objects touch in the model, only one object is identified in the images produced by the GVSPM algorithm and H-GVSPM algorithm. The reconstructed images obtained by the G-CNN algorithm and HG-CNN algorithm not only contain two objects, but also the position of the two objects is closer to the actual position. After processing of the Hadamard product, the artifacts of the reconstructed images obtained using the GH-CNN and HGH-CNN algorithms are significantly fewer than those produced by the G-CNN and HG-CNN algorithms.

### *3.4. The quality evaluation of reconstructed images for lung cross-section models*

In order to reasonably assess the quality of reconstructed images, the correlation coefficient is applied as the evaluation index in this research [24, 32]. If the correlation coefficient is higher, the quality of the reconstructed images is better. On the contrary, the quality of the reconstructed images is poorer. The expression of the correlation coefficient is as follows:

$$ICC = \frac{\sum_{i=1}^{n}(x_i - \bar{x})(y_i - \bar{y})}{\sqrt{\sum_{i=1}^{n}(x_i - \bar{x})^2}\sqrt{\sum_{i=1}^{n}(y_i - \bar{y})^2}} \tag{2}$$

here, *ICC* represents the correlation coefficient. xi and yi represent the conductivity values of the *i*-th finite element between the actual conductivity distribution and the reconstructed conductivity distribution, respectively, *n* represents the total number of the finite elements in the field. $\bar{x}$ and $\bar{y}$ represent the average value of the actual conductivity and the reconstructed conductivity. The correlation coefficients are adopted to evaluate the quality of the reconstructed images for all algorithms.

Table 1. Correlation coefficients of the lung cross-section models

| *ICC* | Model I | Model II | Model III | Model IV | Model V | Model VI | Model VII | Model VIII | Model IX | Model X |
|---|---|---|---|---|---|---|---|---|---|---|
| GVSPM | 0.428 | 0.435 | 0.417 | 0.489 | 0.853 | 0.360 | 0.714 | 0.803 | 0.526 | 0.394 |
| H-GVSPM | 0.468 | 0.411 | 0.472 | 0.509 | 0.806 | 0.414 | 0.740 | 0.829 | 0.570 | 0.362 |
| G-CNN | 0.742 | 0.928 | 0.574 | 0.834 | 0.979 | 0.907 | 0.759 | 0.972 | 0.767 | 0.858 |
| HG-CNN | 0.645 | 0.904 | 0.711 | 0.782 | 0.962 | 0.544 | 0.881 | 0.970 | 0.710 | 0.754 |
| GH-CNN | 0.831 | 0.933 | 0.627 | 0.912 | 0.946 | 0.902 | 0.703 | 0.972 | 0.802 | 0.917 |
| HGH-CNN | 0.735 | 0.900 | 0.797 | 0.815 | 0.918 | 0.547 | 0.781 | 0.928 | 0.769 | 0.749 |

The quality of the reconstructed images for ten types of lung cross-section models is evaluated. The correlation coefficients are presented in Table 1. For Model V and Model VIII, the GVSPM algorithm and H-GVSPM algorithm exhibit the best reconstruction effect, and the correlation coefficient is greater than 0.8. However, the correlation coefficients are relatively small in other models. The correlation coefficients of the G-CNN algorithm and HG-CNN algorithm are greater than those of the GVSPM algorithm and H-GVSPM algorithm, respectively. Therefore, the G-CNN algorithm and HG-CNN algorithm are able to improve the quality of reconstructed images after post-processing of the 1D-CNN. If defined:

$$Rg = \frac{I_{G-CNN}}{I_{GVSPM}}, Rhg = \frac{I_{HG-CNN}}{I_{H-GVSPM}} \tag{3}$$

here $I_{G-CNN}$ represents the correlation coefficient of the G-CNN algorithm, and $I_{HG-CNN}$ represents the correlation coefficient of the HG-CNN algorithm. $I_{GVSPM}$ represents the correlation coefficient of the reconstructed images obtained by the GVSPM algorithm. $I_{H-GVSPM}$ represents the correlation coefficient of the H-GVSPM algorithm.

A conclusion is drawn from Table 1 that the maximum value of *Rg* is 2.519, and the minimum value is 1.063. The maximum value of *Rhg* is 2.200, and the minimum value is 1.170. It shows that the quality of reconstructed images is improved by applying the 1D-CNN to optimize the algorithm. It is known that the correlation coefficients of the GH-CNN algorithm and the HGH-CNN algorithm are also relatively high, as shown in Table 1. The reconstruction effect of the two algorithms is superior to that of the GVSPM algorithm and the H-GVSPM algorithm. In general, it is slightly worse than the G-CNN algorithm and the HG-CNN algorithm. However, the effect is excellent from the perspective of removing artifacts.

## 4. Experimental Analysis

### *4.1. Experimental system and experimental conditions*

The experimental system is an embedded 16-electrode system based on the Zynq processor in this study. The system is composed of a signal acquisition controller, signal excitation module, signal acquisition module, analog multiplexer, and other components. The Zynq-based signal acquisition controller converts digital signals into analog signals using a digital-to-analog converter (DAC). The signal is transmitted to the voltage-controlled current source (VCCS). Here, the high-speed operational amplifier amplifies the excitation signal. In order to ensure that the excitation channel or measurement channel of the analog multiplexer operates on different electrode pairs, the GPIO gating control of the signal acquisition controller enables the switching of the analog multiplexer channel. After the measured boundary voltage is operated by the buffer circuit, filter circuit, and drive current, the continuous analog signal is converted into a discrete digital signal through an analog-to-digital converter (ADC). The discrete digital signal is fed back to the signal acquisition controller. The signal acquisition controller and the PC terminal are connected via the Internet. The ID of the signal acquisition controller is entered into the PC terminal to acquire the discrete signal. The high-speed operational amplifier needs to have strong anti-noise interference capability. Therefore, it is necessary to consider its parameters.

The experimental subjects are living organisms in the lung cross-section models. Sixteen flexible electrodes are attached to the outside of the chests of volunteers. Then these volunteers breathe freely. Next, the boundary voltages are extracted during breathing. The inverse problem of EIT is solved by utilizing these boundary voltages. The structure of the experimental device is shown in Fig. 6. The VCCS is a standard Howland current source with the advantages of fast response speed and high constant current accuracy. There are five resistors R1-R5. Each resistor has a resistance value $R_i$ ($i$=1,2,3,4,5). And there is a load. The operational amplifier circuit comprises a negative feedback circuit and a positive feedback circuit. The negative feedback circuit consists of resistor R2, while the positive feedback circuit comprises resistors R4 and R5.

As can be seen from Fig. 6(b), the load is grounded. The current of the resistor R2 is the same as that of R1. In the meantime, the currents flowing through resistors R4 and R3 are the same. The current of resistance R5 is equal to the sum of the currents between resistance R4 and $i_1$. Finally, the output current is determined by (4) under the condition $(R_5+R_4)/R_3=R_2/R_1=1$.

$$i_1 = \frac{u_1 - u_0}{R_5} \tag{4}$$

## *4.2. Experimental results of lung cross-section models*

Since there is no open-source database of EIT imaging for lung cross-section models. It is not possible to train neural networks with large datasets. More importantly, it is inappropriate to apply the lung cross-section models in simulations. Considering the aforementioned factors, a hypothetical model of the lung is developed in this research, as illustrated in Fig.7. Although the quality of the images reconstructed by the GVSPM algorithm is not optimal, it still maintains a certain level of fidelity. The position and size of the object are able to be roughly evaluated by the reconstructed image. Therefore, the GVSPM algorithm is applied to reconstruct images of the lung cross-section models. The lung outline is approximately delineated based on the conductivity distribution of the reconstructed images. Secondly, the unnecessary elements of the edge are removed. These elements are set as the background element. Next, except for the lung outline, the rest of the elements are set as default background elements. Finally, the ideal model of pulmonary expiratory volume is obtained, and the ideal conductivity distribution is extracted. The optimal method based on the 1D-CNN is utilized to reconstruct the experimental model for the lung cross-section models.

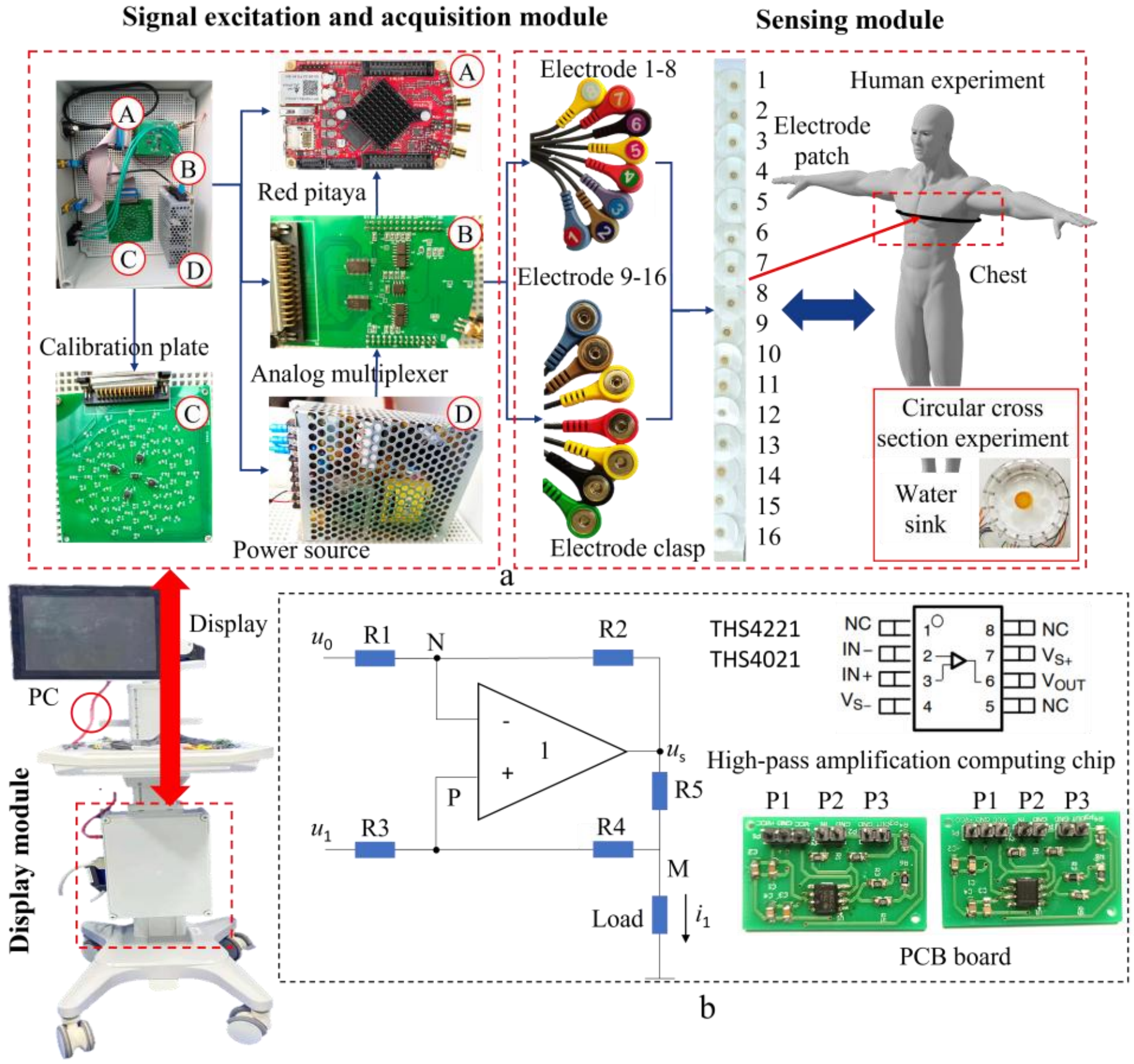

Fig. 6. The structure of the experimental device: (a) the components of the device, (b) the voltage-controlled current source (VCCS) and PCB board.

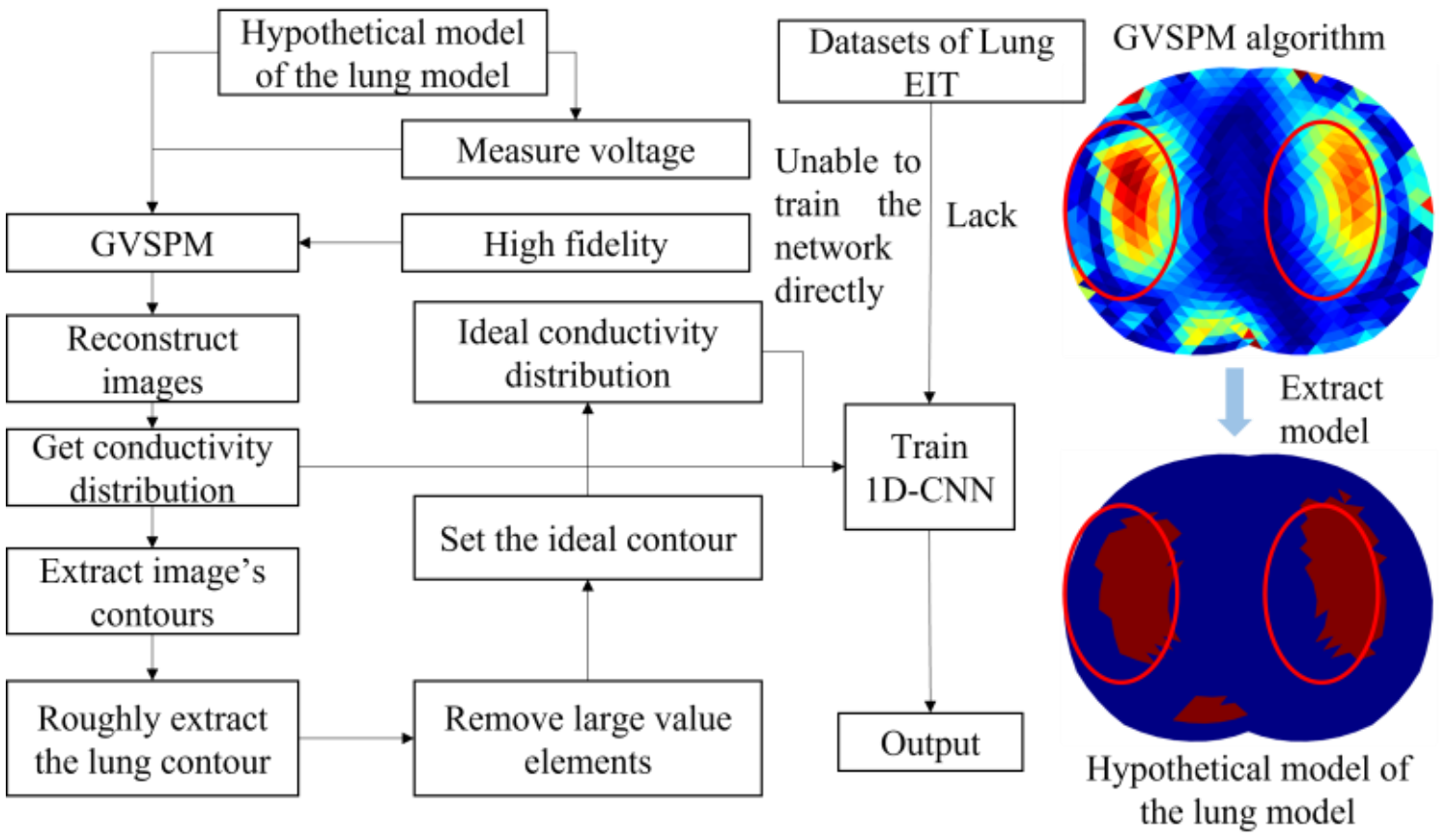

Fig. 7. The hypothetical model of the lung cross-section models

In the simulation results of lung cross-section models, the maximum number of iterations for training the neural network is set to one hundred. Due to the small datasets in the lung experimental model, the maximum number of iterations for training the neural network is set to one thousand. Based on the above assumptions, five hypothetical models are randomly extracted for EIT imaging in this study. The results are shown in Fig.8.

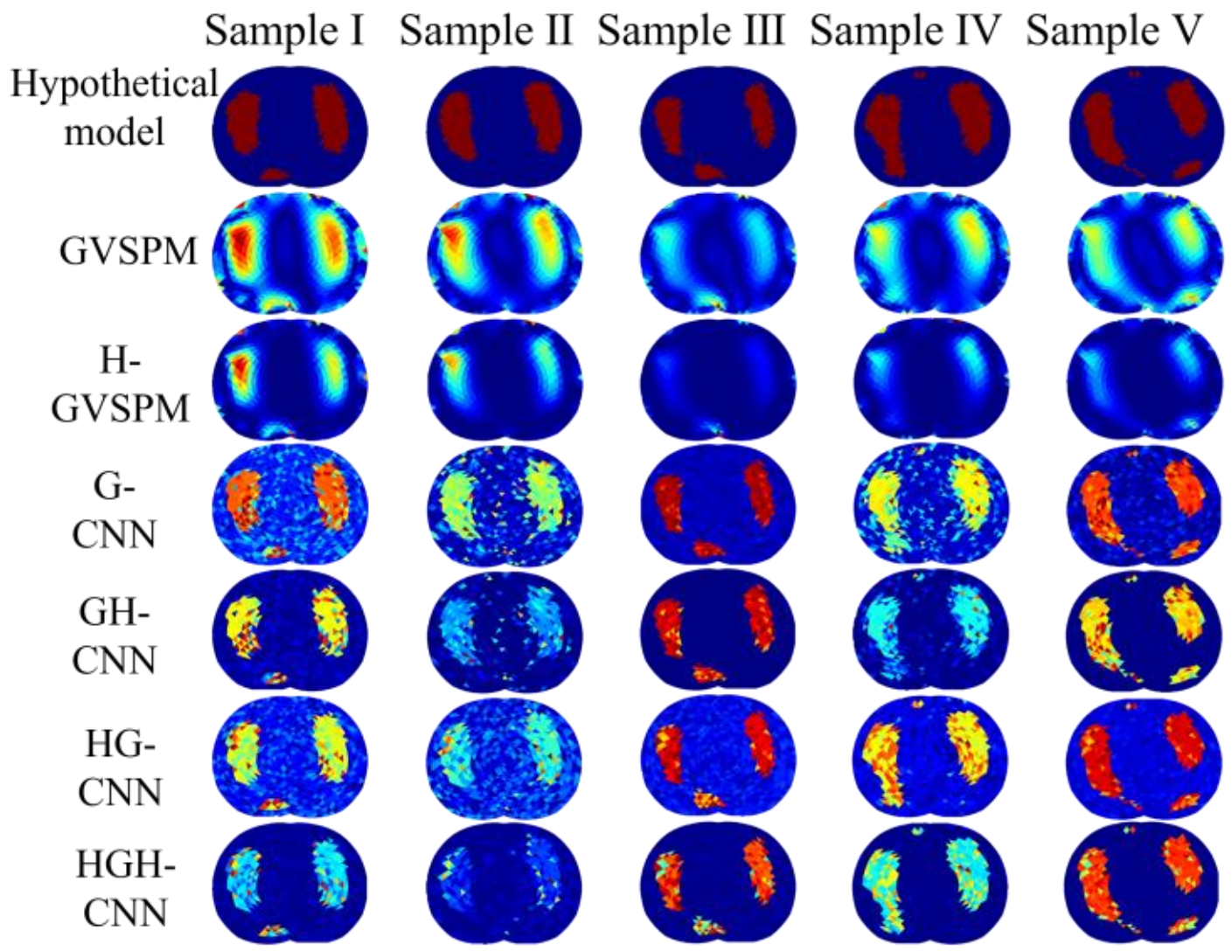

Fig. 8. The experimental results of the hypothetical models

Compared with the GVSPM algorithm, the Hadamard product reduces artifacts and weakens target objects for the H-GVSPM algorithm. The lung contour of the reconstructed images obtained by the G-CNN algorithm is clearer than that obtained by the GVSPM algorithm. The GH-CNN algorithm retains the information of the G-CNN algorithm and reduces artifacts. The reconstruction effect of the HG-CNN algorithm is significantly better than that of the H-GVSPM algorithm, especially in Sample III. The H-GVSPM algorithm produces an image with an unclear contour, whereas the HG-CNN algorithm obtains a clearer lung contour. Intuitively, the reconstruction effect of the HGH-CNN algorithm is better than that of the HG-CNN algorithm. The reconstructed images have fewer artifacts and high quality. In the reconstructed images of Sample III and Sample V, the image quality improved after optimizing the 1D-CNN. What's more, not only does it clearly distinguish lung contours, but makes the image have fewer artifacts and high resolution.

## 5. Conclusion

In order to enhance the resolution and quality of reconstructed images for lung EIT imaging. The G-CNN algorithm and HG-CNN algorithm are proposed based on a one-dimensional convolutional neural network. At the same time, the GH-CNN algorithm and HGH-CNN algorithm are proposed based on the Hadamard product in this research. The aim is to solve the problem that the quality of the reconstructed images is poor for the GVSPM algorithm and H-GVSPM algorithm. It provides a further basis for the feasibility of using EIT imaging to detect lung function diseases. The lung cross-section models are constructed to establish the datasets through simulation. Then, the EIT images of the lung cross-section models are reconstructed. In the simulation results of the lung cross-section models, the correlation coefficients are increased to 2.52 times and 2.20 times for the GVSPM algorithm and H-GVSPM algorithm, respectively. Although the correlation coefficients of the GH-CNN algorithm and HGH-CNN algorithm are not higher than those of the G-CNN algorithm and HG-CNN algorithm, most information is retained. The artifacts are reduced by using the Hadamard product in the reconstructed images. By analyzing the results of five lung-model experiments, it is found that the reconstruction effect of the G-CNN algorithm and HG-CNN algorithm is excellent in all samples. The contour of the objects is more clearly distinguished. The GH-CNN algorithm and HGH-CNN algorithm effectively eliminate most artifacts of the reconstructed images. Therefore, the resolution of the reconstructed images is relatively high. In summary, the G-CNN algorithm and HG-CNN algorithm effectively enhance the reconstruction effect of the GVSPM algorithm and the H-GVSPM algorithm, respectively. The number and contour of the object are well-identified. The imaging effect of the hypothesis model verified that the G-CNN algorithm and HG-CNN algorithm are able to improve the quality of lung EIT imaging for the lung cross-section models. The GH-CNN algorithm and HGH-CNN algorithm effectively remove artifacts and further enhance the quality.

## Acknowledgements

This work was supported in part by the Key Science and Technology Plan Project of Jinhua City, China (2023-3-084); Open Project of Key Laboratory of Reliability of Numerical Control Equipment, Ministry of Education (JLU-cncr-202407); Zhejiang Provincial Natural Science Foundation of China (LZ24E050008); National Natural Science Foundation of China (52205075); National Natural Science Foundation of China (52105564).

Zhenzhong Song, Jianping Li, Jianming Wen, Nen Wan, Haolin Tang Jijie Ma, Yili Hu, Yu Zhang, Yingting Wang, and Kang Chen are with the Key Laboratory of Intelligent Operation and Maintenance Technology & Equipment for Urban Rail Transit of Zhejiang Province, the Institute of Precision Machinery and Smart Structure, College of Engineering, Zhejiang Normal University, Jinhua, Zhejiang 321004, China (e-mail: sphil12@zjnu.edu.cn; lijp@zjnu.cn; wjming@zjnu.cn; wannen@zjnu.cn; 1531627334@qq.com; mjj@zjnu.cn; huyili@zjnu.edu.cn; zjnuzy@zjnu.cn; wangyingting@zjnu.edu.cn; kangchen@zjnu.edu.cn). (Corresponding author: Jianping Li; phone: 178-0589-8098; e-mail: lijp@zjnu.cn).

Jiafeng Yao is with the College of Mechanical and Electrical Engineering, Nanjing University of Aeronautics and Astronautics, Nanjing, 210016 China (e-mail: jiaf.yao@nuaa.edu.cn).

Zengfeng Gao is with the Division of Fundamental Engineering, Department of Mechanical Engineering, Graduate School of Science and Engineering, Chiba University, Chiba 263-8522, Japan (e-mail: gaozengfeng@chiba-u.jp).